\documentstyle[preprint,aps,epsf,floats]{revtex}

\begin{document}
\tighten

\newcommand{\gsim}{\raisebox{-0.7ex}{$\stackrel{\textstyle >}{\sim}$ }}
\newcommand{\lsim}{\raisebox{-0.7ex}{$\stackrel{\textstyle <}{\sim}$ }}
\def\Dslash{D\hskip-0.65em /}
\def\Se{s}
\def\Tr{t}
\def\SD{x}
\def\cb{{\cal B}}
\def\cbb{{\overline{\cal B}}}

\def\Journal#1#2#3#4{{#1} {\bf #2}, #3 (#4)}

\def\NCA{\em Nuovo Cimento}
\def\NIM{\em Nucl. Instrum. Methods}
\def\NIMA{{\em Nucl. Instrum. Methods} A}
\def\NPB{{\em Nucl. Phys.} B}
\def\NPA{{\em Nucl. Phys.} A}
\def\NP{{\em Nucl. Phys.} }
\def\PLB{{\em Phys. Lett.} B}
\def\PRL{\em Phys. Rev. Lett.}
\def\PRD{{\em Phys. Rev.} D}
\def\PRC{{\em Phys. Rev.} C}
\def\PRA{{\em Phys. Rev.} A}
\def\PR{{\em Phys. Rev.} }
\def\ZPC{{\em Z. Phys.} C}
\def\SJP{{\em Sov. Phys. JETP}}
\def\SJNP{{\em Sov. Phys. Nucl. Phys.}}

\def\FBS{{\em Few Body Systems Suppl.}}
\def\IJMP{{\em Int. J. Mod. Phys.} A}
\def\UJP{{\em Ukr. J. of Phys.}}
\def\CJP{{\em Can. J. Phys.}}
\def\SCI{{\em Science} }
\def\AST{{\em Astrophys. Jour.} }

\preprint{\vbox{
\hbox{NT@UW-01-021}
\hbox{UMPP\# 02-006}
}}
\bigskip
\bigskip

\title{Matrix Elements of Twist-2 Operators in Quenched 
Chiral Perturbation  Theory
}
\author{{\bf Jiunn-Wei Chen}$^a$ and {\bf Martin J. Savage}$^{b}$}
\address{$^a$ Department of Physics, University of Maryland, \\
College Park, MD 20742-4111.}
\address{$^b$ Department of Physics, University of Washington, \\
Seattle, WA 98195. }
\maketitle

\begin{abstract}
We compute the leading quark mass dependence of the matrix
elements of isovector twist-2 operators between octet baryon states
in quenched QCD using quenched chiral perturbation theory.
There are contributions of the form $m_q\log m_q$ and $m_q$, 
in analogy with QCD, but
there are also contributions of the form $\log m_q$ 
from hairpin interactions.
The nucleon does not receive such hairpin contributions.
\end{abstract}

\vskip 2in

\vfill\eject

\section{Introduction}

The parton distribution functions (PDF's) 
of the nucleon are fundamental quantities 
associated with the strong interactions.
Extensive experimental investigations have been undertaken during the 
past three decades to measure these distributions via deep-inelastic 
scattering (DIS) of leptons from protons and light nuclei.
Due to the intrinsically non-perturbative nature of the strong interactions
in the low-momentum region,
theoretical efforts to understand these distributions have had only 
limited success.
With the ever increasing power of computers and significant developments in 
algorithms used to numerically simulate QCD on the lattice, it is 
hoped that properties of the PDF's can be determined from
first principle~\cite{latticeQCD,latticeQCDu} at
some point in the not so distant future (for a review see Ref.~\cite{Negele}).
It is the forward matrix elements of twist-2 operators that are
computed numerically and these matrix elements 
are directly related to moments of the PDF's.
Of course, only matrix elements computed with unquenched QCD 
with the physical
values of the quark masses, $m_q$,
are to be directly compared with experimental data, but at
this point in time such computations are not possible.
All present computations are performed with lattice quark masses, 
$m_q^{\rm  latt.}$, that give a pion mass of 
$m^{\rm  latt.}_\pi\sim 500~{\rm MeV}$,
and most simulations are quenched.  
Despite the fact that quenched computations require significantly less
computer time, they cannot, unfortunately, be connected to QCD in any way.
While solid progress is being made toward unquenched 
calculations~\cite{latticeQCDu}
of the moments of the PDF's, 
it is likely that partially-quenched~\cite{Pqqcd} computations will 
first provide a reliable connection between lattice computations
and nature by allowing for calculations with smaller $m_q$ and thereby
minimizing the impact of the $m_q$-extrapolation.
However, it is of interest, from a theoretical standpoint, to know the 
moments of the PDF's in quenched QCD (QQCD).
To extrapolate from $m_q^{\rm  latt.}$ down to $m_q$, 
the $m_q$-dependence 
of the matrix elements is required~\cite{mqothers},
and recently 
chiral perturbation theory ($\chi$PT) has been used to determine
the leading $m_q$-dependence in an expansion about the chiral
limit in QCD~\cite{AS,CJ} and large-$N_c$ QCD~\cite{largeN}.
$\chi$PT also allows for a connection to be made with existing convolution
models of PDF's, and in addition, shows how to make them consistent with 
QCD~\cite{conv}.
In this work we determine the leading $m_q$-dependence of the matrix elements
of isovector twist-2 operators about the chiral limit in QQCD
using quenched chiral perturbation theory 
(Q$\chi$PT)~\cite{Sharpe90,S92,BG92,LS96,Sav01}.

\section{Q$\chi$PT}

The lagrange density of QQCD is
\begin{eqnarray}
{\cal L} & = & 
\sum_{a,b=u,d,s} \overline{q}^a\ 
\left[\ i\Dslash -m_{q}\ \right]_a^b\ q_b
\ +\ 
\sum_{\tilde a,\tilde b=\tilde u, \tilde d,\tilde s}
\overline{\tilde q}^{\tilde a}
\ \left[\ i\Dslash -m_{\tilde q}\ 
\right]_{\tilde a}^{\tilde b}\tilde q_{\tilde b}
\nonumber\\
& = & 
\sum_{j,k=u,d,s,\tilde u, \tilde d,\tilde s} 
\overline{Q}^j\ 
\left[\ i\Dslash -m_{Q}\ \right]_j^k\ Q_k
\ \ \ ,
\label{eq:QQCD}
\end{eqnarray}
where $q$ are the three light-quarks, $u$, $d$, and $s$, and 
$\tilde q$ are three light bosonic quarks  $\tilde u$, $\tilde d$, 
and $\tilde s$.
The super-quark field, $Q_j$, is a six-component column vector with the 
three light-quarks, $u$, $d$, and $s$,
in the upper three entries and the three ghost-light-quarks, 
$\tilde u$, $\tilde d$, and $\tilde s$,
in the lower three entries.
The graded equal-time commutation relations for two fields is
\begin{eqnarray}
Q_i^\alpha ({\bf x}) Q_j^{\beta \dagger} ({\bf y}) - 
(-)^{\eta_i\eta_j}Q_j^{\beta \dagger} ({\bf y})Q_i^\alpha ({\bf x})
& = & 
\delta^{\alpha\beta}\delta_{ij}\delta^3({\bf x}-{\bf y})
\ \ \ ,
\label{eq:comm}
\end{eqnarray}
where $\alpha,\beta$ are spin-indices and $i,j$ are flavor indices.
The objects $\eta_k$ correspond to the parity of the component of $Q_k$,
with $\eta_k=+1$ for $k=1,2,3$ and $\eta_k=0$ for $k=4,5,6$.
The diagonal super mass-matrix, $m_Q$, has entries
$m_Q = {\rm diag}(m_u,m_d,m_s,m_u,m_d,m_s)$,
i.e.  
$m_{\tilde u}=m_u$, $m_{\tilde d}=m_d$ and 
$m_{\tilde s}=m_s$, so that the contribution 
to the determinant in the path integral 
from the $q$'s and the $\tilde q$'s exactly cancel.

In the absence of quark masses,
the lagrange density in eq.~(\ref{eq:QQCD})
has a
graded symmetry $U(3|3)_L\otimes U(3|3)_R$, where the left- and 
right-handed quark fields transform as
$Q_L\rightarrow U_L Q_L$ and $Q_R\rightarrow U_R Q_R$ respectively.
However, the functional integral associated with this Lagrange density 
does not converge unless the transformations on the left- and right-handed
fields are related, ${\rm sdet}(U_L)={\rm sdet}(U_R)$, 
where ${\rm sdet}()$ denotes a superdeterminant~\cite{Pqqcd,S92,BG92},
leaving the theory to have a symmetry
$\left[SU(3|3)_L\otimes SU(3|3)_R\right]\times U(1)_V$, 
where the ``$\times$'' denotes  a semi-direct product as opposed 
to a direct product, ``$\otimes$''.
It is assumed that this symmetry is spontaneously broken 
$\left[SU(3|3)_L\otimes SU(3|3)_R\right]\times U(1)_V\rightarrow 
SU(3|3)_V\times U(1)_V$ so that an identification with QCD can be made.

The pseudo-Goldstone bosons of QQCD form a $6\times 6$ matrix, $\Phi$, 
that can be written in block form
\begin{eqnarray}
\Phi & = & \left(\matrix{\pi & \chi^\dagger \cr \chi & \tilde\pi}\right)
\ \ \ ,
\label{eq:mesondef}
\end{eqnarray}
where $\pi$ is the $3\times 3$ matrix of pseudo-Goldstone bosons 
including the $\eta^\prime$
with quantum numbers of $\overline{q}q$ pairs,
$\tilde \pi$ is a  $3\times 3$ matrix of pseudo-Goldstone bosons 
including the $\tilde\eta^\prime$
with quantum numbers of $\overline{\tilde q}\tilde q$ pairs, and 
$\chi$ is a $3\times 3$ matrix of pseudo-Goldstone fermions
with quantum numbers of $\overline{\tilde q}q$ pairs,
\begin{eqnarray}
\pi & = & \left(\matrix{\eta_u &\pi^+ &K^+ \cr \pi^- &\eta_d & K^0\cr
K^- &\overline{K}^0 & \eta_s}\right)
\ \ ,\ \ 
\tilde\pi \ = \ \left(\matrix{\tilde\eta_u &\tilde\pi^+ &
\tilde K^+ \cr \tilde\pi^- &\tilde\eta_d & \tilde K^0\cr
\tilde K^- &\overline{\tilde K}^0 & \tilde\eta_s}\right)
\ \ ,\ \ 
\chi\ =\  \left(\matrix{\chi_{\eta_u} &\chi_{\pi^+} &\chi_{K^+} \cr 
\chi_{\pi^-} &\chi_{\eta_d} & \chi_{K^0}\cr
\chi_{K^-} &\chi_{\overline{K}^0} &\chi_{\eta_s}}\right)
\ \ ,
\label{eq:mesmats}
\end{eqnarray}
As the object 
\begin{eqnarray}
\Phi_0 & = & {1\over\sqrt{6}}{\rm str}\left(\Phi\right)
\ =\ {1\over\sqrt{2}}\left(\ \eta^\prime - \tilde\eta^\prime\ \right)
\ \ ,
\label{eq:phi0}
\end{eqnarray}
is invariant under 
$\left[SU(3|3)_L\otimes SU(3|3)_R\right]\times U(1)_V$
the most general lagrange density that describes low-momentum 
dynamics will contain arbitrary functions of 
$\Phi_0$~\cite{S92,BG92}.
At lowest order in the chiral expansion, the Lagrange density
that describes the dynamics of the pseudo-Goldstone bosons 
is, using the notation of Ref.~\cite{LS96},
\begin{eqnarray}
{\cal L} & = & 
{f^2\over 8}\ {\rm str}\left[\  
\partial^\mu\Sigma^\dagger\partial_\mu\Sigma\ \right]
\ +\ 
\lambda\ {\rm str} \left[\ m_Q\Sigma\ +\ m_Q^\dagger\Sigma^\dagger\ \right]
\ +\ \alpha_\Phi \ \partial^\mu\Phi_0\partial_\mu\Phi_0
\ -\ 
M_0^2\  \Phi_0^2
\ \ \ ,
\label{eq:lagpi}
\end{eqnarray}
where the parameter $\lambda$ is chosen to reproduce the meson masses,
and $\Sigma$ is the exponential of the $\Phi$ field,
\begin{eqnarray}
\Sigma & = & {\rm exp}\left( {2\  i\  \Phi\over f}\right)
\ \ .
\label{eq:sigdef}
\end{eqnarray}
With this normalization, $f\sim 132~{\rm MeV}$ in QCD.
In addition, 
it is understood that the operators with coefficients $\alpha_\Phi$ and
$M_0^2$, the hairpin interactions,
are inserted perturbatively.
Expanding out the Lagrange density in eq.~(\ref{eq:lagpi}) to quadratic order
in the meson fields, one finds relations between the meson masses in the
isospin
limit,
\begin{eqnarray}
m_{\eta_s}^2 & = & 2 m_K^2 - m_\pi^2
\ \ ,\ \ 
m_{\eta_u}^2 \ = \ m_{\eta_d}^2\ =\ m_\pi^2
\ \ \ .
\label{eq:massrels}
\end{eqnarray}
The Lagrange density in eq.~(\ref{eq:lagpi})
has been used to compute several observables in the
meson sector, such as $f_K$, $f_\pi$~\cite{S91} 
and the meson masses, to one-loop in
perturbation theory~\cite{S92,BG92,Pall97} 
(for a review see Ref.~\cite{G94}).

The inclusion of the lowest-lying baryons, 
the octet of spin-${1\over 2}$ baryons 
and the decuplet of spin-${3\over 2}$ baryon resonances,
is detailed in Ref.~\cite{LS96}.
An interpolating field that has non-zero overlap with the 
baryon octet (when the $ijk$ indices are restricted to $1,2,3$) 
is~\cite{LS96}
\begin{eqnarray}
{\cal B}^\gamma_{ijk} & \sim &
\left[\ Q_i^{\alpha,a} Q_j^{\beta,b} Q_k^{\gamma,c}
\ -\  Q_i^{\alpha,a} Q_j^{\gamma,c} Q_k^{\beta,b}\ \right]
\epsilon_{abc} \left(C\gamma_5\right)_{\alpha\beta}
\ \ \ ,
\label{eq:octinter}
\end{eqnarray}
where $C$ is the charge conjugation operator,
$a,b,c$ are color indices and $\alpha,\beta,\gamma$ are spin indices.
Dropping the spin index, one finds that
under the interchange of flavor indices~\cite{LS96},
\begin{eqnarray}
{\cal B}_{ijk} & = & (-)^{1+\eta_j \eta_k}\  {\cal B}_{ikj}
\ \ ,\ \ 
{\cal B}_{ijk} \ +\  (-)^{1+\eta_i \eta_j}\ {\cal B}_{jik}
\ +\ (-)^{1 + \eta_i\eta_j + \eta_j\eta_k + \eta_k\eta_i}\ 
{\cal B}_{kji}\ =\ 0
\ \ \ .
\label{eq:bianchi}
\end{eqnarray}
The object ${\cal B}_{ijk}$ describes a {\bf 70} dimensional representation
of $SU(3|3)_V$~\cite{LS96}.
It is convenient to 
decompose the irreducible representations of $SU(3|3)_V$ 
into  irreducible representations
$SU(3)_{q}\otimes SU(3)_{\tilde q}\otimes U(1)$~\cite{Sav01}.
The subscript denotes where the ``$SU(3)$'' 
acts, either on the $q$'s or on the
$\tilde q$'s.
The ground floor of the {\bf 70}-dimensional representation contains baryons
that are comprised of three quarks, $qqq$, and transforms as an
$({\bf 8},{\bf 1})$ of  $SU(3)_{q}\otimes SU(3)_{\tilde q}$.
The octet baryons are embedded as
\begin{eqnarray}
{\cal B}_{abc} & = & {1\over\sqrt{6}}
\left( \ \epsilon_{abd}\ B^d_c\ +\ 
\epsilon_{acd} B^d_b\ \right)
\ \ \ ,
\end{eqnarray}
where the indices are restricted to take the values $a,b,c=1,2,3$ only.
The octet baryon matrix is
\begin{eqnarray}
B & = & \left(\matrix{{1\over\sqrt{6}}\Lambda + {1\over\sqrt{2}}\Sigma^0
& \Sigma^+ & p\cr
\Sigma^- & {1\over\sqrt{6}}\Lambda - {1\over\sqrt{2}}\Sigma^0 & n\cr
\Xi^- & \Xi^0 & -{2\over\sqrt{6}}\Lambda}\right)
\ \ \ .
\label{eq:baryons}
\end{eqnarray}
The first floor of the  {\bf 70}-dimensional representation contains baryons
that are composed of two quarks and one ghost-quark, $\tilde q qq$,
and transforms as an
$({\bf 6},{\bf 3})\oplus (\overline{\bf 3},{\bf 3})$
of  $SU(3)_{q}\otimes SU(3)_{\tilde q}$.
The tensor representation $_{\tilde a} \Se_{ab}$ 
of the $({\bf 6},{\bf 3})$ multiplet, where ${\tilde a}=1,2,3$ runs over the 
$\tilde q$ indices and $a,b=1,2,3$ run over the $q$ indices,
has baryon assignment
\begin{eqnarray}
_{\tilde a} \Se_{11} & = & 
\Sigma_{\tilde a}^{+1}
\ \ ,\ \ 
_{\tilde a} \Se_{12}\ =\ _{\tilde a} \Se_{21}\ =\ 
{1\over\sqrt{2}} \Sigma_{\tilde a}^{0}
\ \ ,\ \ 
_{\tilde a} \Se_{22}\ =\ 
\Sigma_{\tilde a}^{-1}
\nonumber\\
_{\tilde a} \Se_{13}& = & _{\tilde a} \Se_{31}
\ = \
{1\over\sqrt{2}}
\ ^{(6)}\Xi_{\tilde a}^{+{1\over 2}}
\ \ ,\ \ 
_{\tilde a} \Se_{23}\ =\ _{\tilde a} \Se_{32}
\ =\ 
{1\over\sqrt{2}}
\ ^{(6)}\Xi_{\tilde a}^{-{1\over 2}}
\ \ ,\ \ 
_{\tilde a} \Se_{33} \ =\ 
\Omega_{\tilde a}^0
\ \ \ .
\label{eq:sixdef}
\end{eqnarray}
The right superscript denotes the third component of q-isospin,
while the left subscript denotes the $\tilde q$ flavor.
The tensor representation $_{\tilde a} \Tr^a$ 
of the $(\overline{\bf 3},{\bf 3})$ 
multiplet, where ${\tilde a}=1,2,3$ runs over the 
$\tilde q$ indices and $a=1,2,3$ run over the $q$ indices,
has baryon assignment
\begin{eqnarray}
_{\tilde a} \Tr^1 & = & ^{(\overline{3})}\Xi_{\tilde a}^{-{1\over 2}}
\ \ ,\ \ 
_{\tilde a} \Tr^2 \ =\  ^{(\overline{3})}\Xi_{\tilde a}^{+{1\over 2}}
\ \ ,\ \ 
_{\tilde a} \Tr^3 \ =\  \Lambda_{\tilde a}^0
\ \ \ .
\label{eq:tripdef}
\end{eqnarray}
The $_{\tilde a} \Se_{ab}$ and the $_{\tilde a} \Tr^a$ 
are uniquely embedded into
${\cal B}_{ijk}$ (up to field redefinitions), 
constrained by the relations in eq.~(\ref{eq:bianchi}):
\begin{eqnarray}
{\cal B}_{ijk} & = & 
\sqrt{2\over 3\ }\ _{i-3}\Se_{jk}
\ \ \ \ \ {\rm for}\  \ \ \ i=4,5,6\ \ {\rm and}\ \ j,k=1,2,3
\nonumber\\
{\cal B}_{ijk} & = & 
{1\over 2}\ \  _{j-3}\Tr^\sigma \varepsilon_{\sigma i k }
\ +\ {1\over\sqrt{6}}\ \ _{j-3}\Se_{ik}
\ \ \ \ \ {\rm for}\  \ \ \ j=4,5,6\ \ {\rm and}\ \ i,k,\sigma =1,2,3
\nonumber\\
{\cal B}_{ijk} & = & 
-{1\over 2}\ \   _{k-3}\Tr^\sigma \varepsilon_{\sigma i j }
\ -\ {1\over\sqrt{6}}\ \  _{k-3}\Se_{ij}
\ \ \ \ \ {\rm for}\  \ \ \ k=4,5,6\ \ {\rm and}\ \ i,j,\sigma =1,2,3
\ \ \ .
\label{eq:firstfloor}
\end{eqnarray}
As we are only interested in one-loop contributions to observables with 
$qqq$-baryons in the asymptotic states, we do not 
explicitly show the second and 
third floors of the {\bf 70}.


An interpolating field that contains the spin-${3\over 2}$
decuplet as the ground floor is~\cite{LS96}
\begin{eqnarray}
{\cal T}^{\alpha ,\mu}_{ijk} & \sim &
\left[
Q^{\alpha,a}_i Q^{\beta,b}_j Q^{\gamma,c}_k +
Q^{\beta,b}_i Q^{\gamma,c}_j Q^{\alpha,a}_k  +
Q^{\gamma,c}_i Q^{\alpha,a}_j Q^{\beta,b}_k 
\right]
\varepsilon_{abc} (C\gamma^\mu)_{\beta\gamma}
\ \ \ ,
\label{eq:tdef}
\end{eqnarray}
where the indices $i,j,k$ run from $1$ to $6$.
Neglecting spin indices, one finds that
under the interchange of flavor indices~\cite{LS96}
\begin{eqnarray}
{\cal T}_{ijk} & = & 
(-)^{1+\eta_i\eta_j} {\cal T}_{jik}\ =\ 
(-)^{1+\eta_j\eta_k} {\cal T}_{ikj}
\ \ \ .
\label{eq:ttrans}
\end{eqnarray}
${\cal T}_{ijk}$ describes a ${\bf 38}$ dimensional representation
of $SU(3|3)_V$, which has a ground floor transforming as 
$({\bf 10},{\bf 1})$ under $SU(3)_q\otimes SU(3)_{\tilde q}$ with
\begin{eqnarray}
{\cal T}_{abc} & = & T_{abc}
\ \ \ ,
\label{eq:Tbarys}
\end{eqnarray}
where the indices are restricted to take the values $a,b,c=1,2,3$,
and where $T_{abc}$ is the totally symmetric tensor containing
the decuplet of baryon resonances,
\begin{eqnarray}
T_{111} & = & \Delta^{++}
\ \ ,\ \ 
T_{112} \ =\  {1\over\sqrt{3}}\Delta^+
\ \ ,\ \ 
T_{122} \ =\  {1\over\sqrt{3}}\Delta^0
\ \ ,\ \ 
T_{222} \ =\   \Delta^{-}
\nonumber\\
T_{113} & = & {1\over\sqrt{3}}\Sigma^{*,+}
\ \ ,\ \ 
T_{123} \ =\  {1\over\sqrt{6}}\Sigma^{*,0}
\ \ ,\ \ 
T_{223} \ =\  {1\over\sqrt{3}}\Sigma^{*,-}
\nonumber\\
T_{133} & = & {1\over\sqrt{3}}\Xi^{*,0}
\ \ ,\ \ 
T_{233} \ =\  {1\over\sqrt{3}}\Xi^{*,-}
\ \ ,\ \ 
T_{333} \ =\  \Omega^{-}
\ \ \ .
\label{eq:decuplet}
\end{eqnarray}
The first floor transforms as a $({\bf 6},{\bf 3})$ under
$SU(3)_q\otimes SU(3)_{\tilde q}$ and has a tensor representation,
$_{\tilde a} \SD_{ij}$, with baryon assignment
\begin{eqnarray}
_{\tilde a} \SD_{11} & = & 
\Sigma_{\tilde a}^{*,+1}
\ \ ,\ \ 
_{\tilde a} \SD_{12}\ =\ _{\tilde a} \SD_{21}\ =\ 
{1\over\sqrt{2}} \Sigma_{\tilde a}^{*,0}
\ \ ,\ \ 
_{\tilde a} \SD_{22}\ =\ 
\Sigma_{\tilde a}^{*,-1}
\nonumber\\
_{\tilde a} \SD_{13} & = & _{\tilde a} \SD_{31}
\ =\ 
{1\over\sqrt{2}}
\ \Xi_{\tilde a}^{*,+{1\over 2}}
\ \ ,\ \ 
_{\tilde a} \SD_{23}\ =\ _{\tilde a} \SD_{32}
\ =\ 
{1\over\sqrt{2}}
\ \Xi_{\tilde a}^{*,-{1\over 2}}
\ \ ,\ \ 
_{\tilde a} \SD_{33} \ =\  
\Omega_{\tilde a}^{*,0}
\ \ \ .
\label{eq:sixTdef}
\end{eqnarray}
The embedding of $_{\tilde a} \SD_{ij}$ into ${\cal T}_{ijk}$ is 
unique (up to field redefinitions),
constrained by the symmetry properties in eq.~(\ref{eq:ttrans}):
\begin{eqnarray}
{\cal T}_{ijk} & = & 
+ {1\over\sqrt{3}}\ _{i-3}\SD_{jk}
\ \ \ \ \ {\rm for}\  \ \ \ i=4,5,6\ \ {\rm and}\ \ j,k=1,2,3
\nonumber\\
{\cal T}_{ijk} & = & 
-{1\over\sqrt{3}}\ \ _{j-3}\SD_{ik}
\ \ \ \ \ {\rm for}\  \ \ \ j=4,5,6\ \ {\rm and}\ \ i,k =1,2,3
\nonumber\\
{\cal T}_{ijk} & = & 
+ {1\over\sqrt{3}}\ \  _{k-3}\SD_{ij}
\ \ \ \ \ {\rm for}\  \ \ \ k=4,5,6\ \ {\rm and}\ \ i,j =1,2,3
\ \ \ .
\label{eq:firstfloorT}
\end{eqnarray}


The free Lagrange density for the ${\cal B}_{ijk}$ and 
${\cal T}_{ijk}$ fields is~\cite{LS96,Sav01}, 
at leading order in the heavy baryon 
expansion~\cite{JMheavy,JMaxial,Jmass,chiralN,chiralUlf}, 
\begin{eqnarray}
{\cal L} & = & 
i\left(\overline{\cal B} v\cdot {\cal D} {\cal B}\right)
\ +\ 2\alpha_M \left(\overline{\cal B}{\cal B}{\cal M}_+\right)
\ +\ 2\beta_M \left(\overline{\cal B}{\cal M}_+{\cal B}\right)
\ +\ 2\sigma_M \left(\overline{\cal B}{\cal B}\right)\ 
{\rm str}\left({\cal M}_+\right)
\nonumber\\
& - & 
i \left(\overline{\cal T}^\mu v\cdot {\cal D} {\cal T}_\mu\right)
\ +\ 
\Delta\ \left(\overline{\cal T}^\mu {\cal T}_\mu\right)
\ +\ 2\gamma_M \left(\overline{\cal T}^\mu{\cal M}_+{\cal T}_\mu\right)
\ -\ 2 \overline{\sigma}_M  \left(\overline{\cal T}^\mu {\cal T}_\mu\right)\
{\rm str}\left({\cal M}_+\right)
\ \ ,
\label{eq:free}
\end{eqnarray}
where 
${\cal M}_+={1\over 2}\left(\xi^\dagger m_Q\xi^\dagger + \xi m_Q\xi\right)$,
$\Delta$ is the decuplet-octet mass splitting,
$v_\mu$ is the baryon four-velocity,
and $\xi=\sqrt{\Sigma}$.
The brackets, $\left(\ ...\ \right)$ denote contraction of lorentz and flavor
indices as defined in Refs.~\cite{LS96,Sav01}.

The Lagrange density describing the interactions of the baryons with the
pseudo-Goldstone bosons is~\cite{LS96}
\begin{eqnarray}
{\cal L} & = & 
2\alpha\ \left(\overline{\cal B} S^\mu {\cal B} A_\mu\right)
\ +\ 
2\beta\ \left(\overline{\cal B} S^\mu A_\mu {\cal B} \right)
\ +\ 
2\gamma\ \left(\overline{\cal B} S^\mu {\cal B} \right)
\ {\rm str}\left(A_\mu \right)
\nonumber\\
+ & &  
2{\cal H} \left(\overline{\cal T}^\nu S^\mu A_\mu {\cal T}_\nu \right)
\ +\ 
\sqrt{3\over 2}{\cal C} 
\left[\ 
\left( \overline{\cal T}^\nu A_\nu {\cal B}\right)\ +\ 
\left(\overline{\cal B} A_\nu {\cal T}^\nu\right)\ \right]
\ +\ 2\gamma^\prime
\left(\overline{\cal T}^\nu S^\mu {\cal T}_\nu\right) 
\ {\rm str}\left(A_\mu \right)
\ ,
\label{eq:ints}
\end{eqnarray}
where $S^\mu$ is the covariant spin-vector~\cite{JMheavy,JMaxial,Jmass}.
Restricting oneself to the $qqq$ sector, it is straightforward to show
that in QCD
\begin{eqnarray}
\alpha & =& {2\over 3}D+2F
\ \ ,\ \ 
\beta \ =\ -{5\over 3}D + F 
\ \ ,
\label{eq:couplings}
\end{eqnarray}
where $D$ and $F$ are constants that multiply the $SU(3)_q$ invariants
that are commonly used in
QCD, but it should be stressed that the $F$ and $D$ in QQCD
will not have the numerical values of those of QCD.
In QQCD there is an additional coupling that must be considered, 
$\gamma$,  
a hairpin interaction~\cite{LS96},
that is usually not considered in the $\chi$PT description of 
low-energy QCD.
In our calculation of the 
matrix elements of  twist-2 operators we will
replace $\alpha$ and $\beta$ with $F$ and $D$, but we will keep $\gamma$ 
explicit.
In the above discussion, vector and axial-vector meson
fields have been introduced in direct analogy with QCD.
The covariant derivative acting on either the ${\cal B}$ or ${\cal T}$ fields
has the form
\begin{eqnarray}
\left({\cal D}^\mu{\cal B}\right)_{ijk} & = & 
\partial^\mu {\cal B}_{ijk}
+
\left(V^\mu\right)^l_i {\cal B}_{ljk}
+ 
(-)^{\eta_i (\eta_j+\eta_m)} \left(V^\mu\right)^m_j {\cal B}_{imk}
+ (-)^{(\eta_i+\eta_j) (\eta_k+\eta_n)}
\left(V^\mu\right)^n_k {\cal B}_{ijn}
\ ,
\label{eq:covariant}
\end{eqnarray}
where the vector and axial-vector meson fields are
\begin{eqnarray}
V^\mu & = & {1\over 2}\left(\ \xi\partial^\mu\xi^\dagger
\ + \ 
\xi^\dagger\partial^\mu\xi \ \right)
\ \ ,\ \ 
A^\mu \ =\  {i\over 2}\left(\ \xi\partial^\mu\xi^\dagger
\ - \ 
\xi^\dagger\partial^\mu\xi \ \right)
\ \ \ .
\label{eq:mesonfields}
\end{eqnarray}

\section{Isovector  Twist-2 Operators}

In QCD, the nonsinglet twist-2 operators have the form,
\begin{eqnarray}
{\cal O}^{ (n), a}_{\mu_1\mu_2\ ... \mu_n}
& = & 
{1\over n!}\ 
\overline{q}\ \lambda^a\ \gamma_{ \{\mu_1  } 
\left(i \stackrel{\leftrightarrow}{D}_{\mu_2}\right)\ 
... 
\left(i \stackrel{\leftrightarrow}{D}_{ \mu_n\} }\right)\ q
\ -\ {\rm traces}
\ \ \ ,
\label{eq:twistop}
\end{eqnarray}
where the $\{ ... \}$ denotes symmetrization on all Lorentz indices,
and where $\lambda^a$ are Gell-Mann matrices acting in flavor-space.
The ${\cal O}^{ (n), a}_{\mu_1\mu_2\ ... \mu_n}$
transform as $({\bf 8},{\bf 1})\oplus  ({\bf 1},{\bf 8})$
under $SU(3)_L\otimes SU(3)_R$ chiral transformations~\cite{AS,CJ}.
Of particular interest to us are the isovector operators where
$\lambda^3={\rm diag}(1,-1,0)$.
In the isospin limit, 
there are no contributions to the matrix elements of
$\lambda^3$ from disconnected diagrams.
In QQCD the nonsinglet  twist-2 operators have the form
\begin{eqnarray}
^Q{\cal O}^{(n), a}_{\mu_1\mu_2\ ... \mu_n}
& = & 
{1\over n!}\ 
\overline{Q}\ \overline{\lambda}^a\ \gamma_{ \{\mu_1  } 
\left(i \stackrel{\leftrightarrow}{D}_{\mu_2}\right)\ 
... 
\left(i \stackrel{\leftrightarrow}{D}_{ \mu_n\} }\right)\ Q
\ -\ {\rm traces}
\ \ \ ,
\label{eq:Qtwistop}
\end{eqnarray}
where the $\overline{\lambda}^a$ are super Gell-Mann matrices,
and we are interested in the isovector matrix elements with
$\overline{\lambda}^3={\rm diag}(1,-1,0,1,-1,0)$~\footnote{
In principle, 
the most general diagonal, isovector, 
supertraceless charge matrix that could be considered
is $\overline{\lambda}^3={\rm diag}(1,-1,0,x,y,-(x+y))$
(for a  discussion about the extension of electro-weak operators contributing
to nonleptonic decays in QQCD see Refs.~\cite{GP,CSPQ}).
For arbitrary $x$ and $y$ there will be contributions from disconnected ghost
quark diagrams.  However, for $y=-x$ there will be no contributions from
disconnected diagrams involving quarks or ghost-quarks in the isospin limit.
In this work we have chosen to examine the case where $x=-y=+1$.
}.
The operators $^Q{\cal O}^{(n), 3}_{\mu_1\mu_2\ ... \mu_n}$
transform as
$({\bf 8},{\bf 1},{\bf 1},{\bf 1})\oplus ({\bf 1},{\bf 8},{\bf 1},{\bf 1})
\oplus ({\bf 1},{\bf 1},{\bf 8},{\bf 1})
\oplus ({\bf 1},{\bf 1},{\bf 1},{\bf 8})$ 
under $SU(3)_{Lq}\otimes SU(3)_{L\tilde q}\otimes SU(3)_{Rq} 
\otimes SU(3)_{R\tilde q}$ graded chiral transformations.

\subsection{Matrix Elements in the Pion}

At leading order in the chiral expansion, matrix elements of the isovector
operator
$^Q{\cal O}^{(n), 3}_{\mu_1\mu_2\ ... \mu_n}$
between meson states are reproduced by operators of the form~\cite{AS}
\begin{eqnarray}
^Q{\cal O}^{(n),3}_{\mu_1\mu_2 ...\mu_n}
& \rightarrow &
a^{(n)} \left(i\right)^n {f^2\over 4} 
\left({1\over\Lambda_\chi}\right)^{n-1}
\ {\rm str}\left[\ 
\Sigma^\dagger  \overline{\lambda}^3 \overrightarrow\partial_{\mu_1}
 \overrightarrow\partial_{\mu_2}...
 \overrightarrow\partial_{\mu_n}
\Sigma
\ +\ 
\Sigma \overline{\lambda}^3 \overrightarrow\partial_{\mu_1}
 \overrightarrow\partial_{\mu_2}...
 \overrightarrow\partial_{\mu_n}
\Sigma^\dagger \right]
\nonumber\\
& & 
\ -\ {\rm traces}
\ \ \ ,
\label{eq:mesontree}
\end{eqnarray}
where $\Lambda_\chi = 4\pi f$ is the scale of chiral symmetry breaking.
Operators involving more derivatives on the meson fields or 
more insertions of $m_q$ make contributions to the matrix
elements that are higher order in the chiral expansion~\cite{AS}.
\begin{figure}[!ht]
\centerline{{\epsfxsize=2.5in \epsfbox{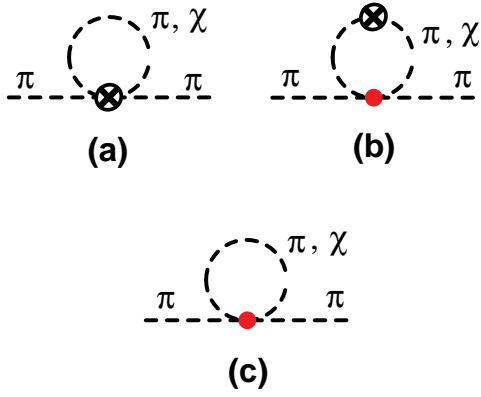}}} 
\vskip 0.15in
\noindent
\caption{\it 
One-loop diagrams that contribute
to the matrix elements of 
$^Q{\cal O}^{(n), 3}_{\mu_1\mu_2\ ... \mu_n}$
in the pion.
The crossed circle denotes an insertion of an operator from
eq.~(\ref{eq:mesontree}), arising directly from the twist-2 operator.
The smaller solid circle denotes an insertion of a leading order 
strong-interaction vertex from eq.~(\ref{eq:lagpi}).
Diagrams (a) and (b) are vertex corrections while diagram (c) denotes
wavefunction renormalization.
}
\label{fig:pidiags}
\vskip .2in
\end{figure}
In QCD, the one-loop diagrams shown in fig.~\ref{fig:pidiags} 
with only the octet mesons in the loop
give rise to the
leading non-analytic terms in $m_q$, of the form $m_q\log m_q$.
Such diagrams have the potential to give analogous contributions to the matrix
elements in QQCD.
However, explicit computation proves that such
non-analytic contributions to the 
matrix element of 
$^Q{\cal O}^{(n), 3}_{\mu_1\mu_2\ ... \mu_n}$
resulting from the diagrams shown in fig.~\ref{fig:pidiags} vanish,
due to a cancellation between the ``$\pi$''-loops and ``$\chi$''-loops.
Therefore, at one-loop order there are no nonanalytic contributions and
the matrix element of 
$^Q{\cal O}^{(n), 3}_{\mu_1\mu_2\ ... \mu_n}$ in the pion is 
\begin{eqnarray}
\langle ^Q{\cal O}^{(n), 3}_{\mu_1\mu_2\ ... \mu_n}\rangle_\pi
& = & 
i \ 4\  a^{(n)} 
\ \left({1\over\Lambda_\chi}\right)^{n-1}
\ \varepsilon^{\alpha\beta 3} 
\ q_{\mu_1} ... q_{\mu_n}
\ -\ {\rm traces} 
\ +\ ...
\ \ \ ,
\label{eq:pitotal}
\end{eqnarray}
for n-odd,
between an initial pion with isospin index $\alpha$ and momentum $q_\mu$, 
and a final pion with isospin index $\beta$.
The matrix elements vanish for n-even and 
the ellipses denote terms that are analytic in $m_q$,
or are higher order in the  chiral expansion.

\subsection{Matrix Elements in the Octet Baryons}

At leading order in the chiral expansion, the matrix elements of
$^Q{\cal O}^{(n), 3}_{\mu_1\mu_2\ ... \mu_n}$
are described by
\begin{eqnarray}
\label{eq:baryonops}
^Q{\cal O}^{(n),3}_{\mu_1\mu_2 ...\mu_n}
& \rightarrow &
\alpha^{(n)}\ v_{\mu_1} v_{\mu_2}...v_{\mu_n}\ 
\left(\ \overline{\cal B}\  {\cal B}\  \overline{\lambda}^3_{\xi +}\ \right)
\ +\ 
\beta^{(n)}\ v_{\mu_1} v_{\mu_2}...v_{\mu_n}\ 
\left(\ \overline{\cal B}\  \overline{\lambda}^3_{\xi +}\  {\cal B}\ \right)
\nonumber\\
& + & 
\gamma^{(n)} 
\ v_{\mu_1} v_{\mu_2}...v_{\mu_n}\ 
\left(\ \overline{\cal T}^\alpha\  \overline{\lambda}^3_{\xi +}
\ {\cal T}_\alpha
\right)
\ +\ 
\sigma^{(n)} {1\over n !}
\ v_{\{ \mu_1} v_{\mu_2}...v_{\mu_{n-2}}\ 
\left(\ \overline{\cal T}_{\mu_{n-1}}\  \overline{\lambda}^3_{\xi +}\ 
{\cal T}_{\mu_n\}}
\right)
\nonumber\\
& & 
\ -\ {\rm traces}
\ ,
\end{eqnarray}
where the $\{... \}$ brackets denote complete symmetrization of the enclosed
indices,
$v_\mu$ is the four-velocity of the heavy baryon, and 
\begin{eqnarray}
\overline{\lambda}^a_{\xi +} & = & 
{1\over 2}
\left(\xi \overline{\lambda}^a \xi^\dagger  
+ \xi^\dagger\overline{\lambda}^a \xi\right)
\ \ \ ,
\end{eqnarray}
in complete analogy with the work of Refs.~\cite{AS,CJ}.
The index contractions denoted by $\left(...\right)$ are the same as 
those in 
eq.~(\ref{eq:ints}) defined in Refs.~\cite{LS96,Sav01}.
In general, the coefficients $\alpha^{(n)},\beta^{(n)},\gamma^{(n)}$ and
$\sigma^{(n)}$ are not constrained by symmetries and
must be determined from elsewhere. However for $n=1$ they are
fixed by the isospin charge of the baryons to be 
\begin{eqnarray}
\alpha^{(1)} & = & +2
\ \ \ ,\ \ \ 
\beta^{(1)}\ =\ +1
\ \ \ ,\ \ \ 
\gamma^{(1)}\ =\ -3
\ \ \ ,\ \ \ 
\sigma^{(1)}\ =\ 0
\ \ \ .
\end{eqnarray}
An expression analogous to eq.~(\ref{eq:baryonops}) exists in QCD, and can be
obtained from eq.~(\ref{eq:baryonops}) by restricting the flavor-indices in
each sum to take values $i=1,2,3$ only. However, one must bear in mind that,
in general,
the values for  $\alpha^{(n)},\beta^{(n)},\gamma^{(n)}$ and $\sigma^{(n)}$
in QQCD will differ from those in QCD for $n\ne 1$.

At next-to-leading order (NLO) there are tree-level 
contributions from local counterterms
involving one insertion of the quark mass matrix.
In QCD we write the operator set as
\begin{eqnarray}
\label{eq:ctQCD}
{\cal O}^{(n),3}_{\mu_1\mu_2 ...\mu_n}
& \rightarrow &
\left(\ 
b_1^{(n)}\ {\rm Tr}\left[\ \overline{B}
\left[\left[\lambda^3_{\xi+} , B\right] ,    m_{q\xi+}\right]\right]
\ +\ 
b_2^{(n)}\ {\rm Tr}\left[\ \overline{B} 
\{ \left[\lambda^3_{\xi+} , B\right] ,    m_{q\xi+} \}\right]
\right.\nonumber\\ & & \left.
+\  b_3^{(n)}\ {\rm Tr}\left[\ \overline{B} \left[\{ \lambda^3_{\xi+} , B\}
 ,    m_{q\xi+}\right]\right]
\ +\ 
b_4^{(n)}\ {\rm Tr}\left[\ \overline{B} \{\{ \lambda^3_{\xi+} , B\}
 ,    m_{q\xi+}\}\right]
\right.\nonumber\\ & & \left.
+ \ b_5^{(n)}\ {\rm Tr}\left[\ \overline{B} B \ \right]\ 
{\rm Tr}\left[\ \lambda^3_{\xi+} m_{q\xi+}\  \right]
\ + \ b_6^{(n)}\ {\rm Tr}\left[\ \overline{B} B \lambda^3_{\xi+} \ \right]\ 
{\rm Tr}\left[\ m_{q\xi+}\  \right]
\right.\nonumber\\ & & \left.
+ \ b_7^{(n)}\ {\rm Tr}\left[\ \overline{B} \lambda^3_{\xi+} B\ \right]\ 
{\rm Tr}\left[\ m_{q\xi+}\  \right]
\ \right)  v_{\mu_1}...v_{\mu_n}\ 
\ -\ {\rm traces}
\ \ \ ,
\end{eqnarray}
where 
$\lambda^3_{\xi+}={1\over 2}\left(\xi\lambda^3\xi^\dagger +
\xi^\dagger\lambda^3\xi\right)$ and 
$m_{q\xi+} = {1\over 2}\left(\xi m_q\xi +
\xi^\dagger m_q \xi^\dagger\right)$.
The $b_i^{(n)}$ must be determined, however, isovector charge conservation
ensures that $b_i^{(1)}=0$.
The explicit renormalization scale dependence of the matrix element at each
order in the chiral expansion requires that, in general, 
the $b_i^{(n)}$ have logarithmic dependence upon the renormalization scale,
$\mu$.
In QQCD the analogous operator set takes the form~\cite{CSPQ}
\begin{eqnarray}
^{Q}{\cal O}^{(n),3}_{\mu_1\mu_2 ...\mu_n}
& \rightarrow &
\left[\ 
b_1^{(n)}\  
\cbb^{kji}\ \{\  \overline{\lambda}^3_{\xi+}\ ,\ {\cal M}_+\ \}^n_i\ 
\cb_{njk}
\right.\nonumber\\ & & \left.
\ +\ 
b_2^{(n)}\ (-)^{(\eta_i+\eta_j)(\eta_k+\eta_n)}\ 
\cbb^{kji}\ \{\  \overline{\lambda}^3_{\xi+}\ ,\ {\cal M}_+\ \}^n_k\ \cb_{ijn}
\right.\nonumber\\ & & \left.
\ +\ 
b_3^{(n)}\  (-)^{\eta_l (\eta_j+\eta_n)}\
\cbb^{kji}\  \left(\overline{\lambda}^3_{\xi+}\right)^l_i\ 
\left( {\cal M}_+\right)^n_j
\cb_{lnk}
\right.\nonumber\\ & & \left.
+\ 
b^{(n)}_4 \  (-)^{\eta_l \eta_j+1}\ 
\cbb^{kji}\ \left(  
\left(\overline{\lambda}^3_{\xi+}\right)^l_i\ \left( {\cal M}_+\right)^n_j
\ +\ \left( {\cal M}_+\right)^l_i 
\left(\overline{\lambda}^3_{\xi+}\right)^n_j \right)
\cb_{nlk}
\right.\nonumber\\ & & \left.
+\ b_5^{(n)}\  (-)^{\eta_i(\eta_l+\eta_j)}\ 
\cbb^{kji} \left(\overline{\lambda}^3_{\xi+}\right)^l_j 
\left( {\cal M}_+\right)^n_i
\cb_{nlk}
\right.\nonumber\\ & & \left.
\ +\ b_6^{(n)}\  \cbb^{kji}  \left(\overline{\lambda}^3_{\xi+}\right)^l_i 
\cb_{ljk}
\ {\rm str}\left( {\cal M}_+ \right) 
\right.\nonumber\\ & & \left.
\ +\ b_7^{(n)}\  \cbb^{kji}  
\left(\overline{\lambda}^3_{\xi+}\right)^n_k \cb_{ijn}
\ {\rm str}\left( {\cal M}_+ \right) 
\ +\ b_8^{(n)}\ \cbb^{kji}\ \cb_{ijk} 
\ {\rm str}\left(\overline{\lambda}^3_{\xi+}\   {\cal M}_+ \right) 
\right.\nonumber\\ & & \left.
\right]\ v_{\mu_1} v_{\mu_2}...v_{\mu_n}\ 
\ -\ {\rm traces}
\ ,
\label{eq:tcts}
\end{eqnarray}
where the $b_i^{(n)}$ of QQCD will, in general, differ from those of QCD.
Isovector charge conservation forces $b_i^{(1)}=0$.

At tree-level the matrix elements in an octet baryon  ``$i$'' is
\begin{eqnarray}
\langle {\cal O}^{(n),3}_{\mu_1\mu_2 ...\mu_n}\rangle_i^{\rm tree}
& = &  v_{\mu_1} v_{\mu_2}...v_{\mu_n}
\left(\ 
\ {\cal M}^{(0)\rm tree}_i\ +\ {\cal M}^{(1)\rm tree}_i\ \right)
\ -\ {\rm traces}
\nonumber\\
\langle ^Q{\cal O}^{(n),3}_{\mu_1\mu_2 ...\mu_n}\rangle_i^{\rm tree}
& = &   
v_{\mu_1} v_{\mu_2}...v_{\mu_n}\ 
\left(\ 
\ ^Q{\cal M}^{(0)\rm tree}_i\ +\ ^Q{\cal M}^{(1)\rm tree}_i\ \right)
\ -\ {\rm trace}
\ \ \ ,
\label{eq:treemats}
\end{eqnarray}
where the Clebsch-Gordan coefficients, ${\cal M}^{\rm tree}_i$ and 
$^Q{\cal M}^{\rm tree}_i$, are given in
Tables~\ref{tab:tree}, \ref{tab:mqQCD} and \ref{tab:mqQQCD}.  
Parametrically, but not numerically,
the leading order matrix elements in QCD and QQCD are the same at
tree-level.
\begin{table}[!ht]
\begin{tabular}[h]{cccc} 
& Baryon & ${\cal M}^{(0) \rm tree}$\  ,\  $^Q{\cal M}^{(0) \rm tree}$ 
& \\ \hline
& $p$ &\ \ ${1\over 3}\left(2\alpha^{(n)}-\beta^{(n)}\right)$& \\
& $\Sigma^+$ &\ \ ${1\over 6}\left(5\alpha^{(n)}+2\beta^{(n)}\right)$& \\
& $\Xi^0$ &\ \ ${1\over 6}\left(\alpha^{(n)}+4\beta^{(n)}\right)$&
\end{tabular}
\caption{
The leading order
tree-level matrix elements of the isovector twist-2 operators
between octet baryon states.
The constants $\alpha^{(n)}$ and $\beta^{(n)}$ are defined in
eq.~(\protect\ref{eq:baryonops}).
}
\label{tab:tree}
\end{table}
\begin{table}[!ht]
\begin{tabular}[h]{cc}
B  & ${\cal M}^{(1) \rm tree}$ (QCD)\\ \hline
$p$ & 
$\overline{m}\  (-b_1^{(n)} + b_2^{(n)}-b_3^{(n)}+b_4^{(n)}+2 b_7^{(n)})
\ +\  m_s\  (b_1^{(n)}+b_2^{(n)}+b_3^{(n)}+b_4^{(n)}+b_7^{(n)})$
\\
$\Sigma^+$ &
$\overline{m}\  (4b_2^{(n)}-2b_6^{(n)}+2 b_7^{(n)})
\ +\  m_s\  (-b_6^{(n)}+b_7^{(n)})$
\\
$\Xi^0$ & 
$\overline{m}\  (b_1^{(n)} + b_2^{(n)}-b_3^{(n)}-b_4^{(n)}-2 b_6^{(n)})
\ +\  m_s\  (-b_1^{(n)}+b_2^{(n)}+b_3^{(n)}-b_4^{(n)}-b_6^{(n)})$
\\
\end{tabular}
\caption{
NLO tree-level contributions to the matrix elements of
${\cal O}^{(n),3}_{\mu_1\mu_2 ...\mu_n}$
in the octet baryons.
$\overline{m}=m_u=m_d$ in the isospin limit.
}
\label{tab:mqQCD}
\end{table}
\begin{table}[!ht]
\begin{tabular}[h]{cc}
B  & $^Q{\cal M}^{(1) \rm tree}$ (QQCD)\\ \hline
$p$ & 
${1\over 3}\  \overline{m}\ 
(-2b_1^{(n)} + 4b_2^{(n)}-b_3^{(n)}+b_4^{(n)}+2 b_5^{(n)})$
\\
$\Sigma^+$ &
$ {1\over 3}\  \overline{m}\ 
(2 b_1^{(n)}+ 5b_2^{(n)}+{1\over 2} b_3^{(n)} + b_4^{(n)}
+ {1\over 2} b_5^{(n)}))
\ +\  {1\over 3}\  m_s
\ ({1\over 2} b_3^{(n)} - 2b_4^{(n)} + 2 b_5^{(n)})$
\\
$\Xi^0$ & 
${1\over 3}\  \overline{m}
\  (4 b_1^{(n)} + b_2^{(n)})
\ +\ {1\over 3}  m_s \ (2 b_3^{(n)}-2 b_4^{(n)}+{1\over 2} b_5^{(n)})$
\\
\end{tabular}
\caption{
NLO tree-level contributions to the matrix elements of
$^Q{\cal O}^{(n),3}_{\mu_1\mu_2 ...\mu_n}$
in the octet baryons.
$\overline{m}=m_u=m_d$ in the isospin limit.
}
\label{tab:mqQQCD}
\end{table}
We consider matrix elements in only the proton, $\Sigma^+$ and $\Xi^0$, as
matrix elements in the other members of the octet are either trivially related
to these, or vanish by isospin symmetry.

In QCD, contributions that are non-analytic in $m_q$ first appear at NLO 
and have the form $m_q\log m_q$~\cite{mqothers}.
However, in QQCD there are additional contributions of the form
$\log m_q$ to baryons other than the nucleon from hairpin interactions.
We will consider the two distinct contributions separately, and write the 
matrix element at one-loop level as 
\begin{eqnarray}
\langle {\cal O}^{(n),3}_{\mu_1\mu_2 ...\mu_n}\rangle_i^
& = & 
\langle {\cal O}^{(n),3}_{\mu_1\mu_2 ...\mu_n}\rangle_i^{\rm tree}
\ +\ 
\langle {\cal O}^{(n),3}_{\mu_1\mu_2 ...\mu_n}\rangle_i^{\rm non-HP}
\nonumber\\
\langle ^Q{\cal O}^{(n),3}_{\mu_1\mu_2 ...\mu_n}\rangle_i^
& = & 
\langle ^Q{\cal O}^{(n),3}_{\mu_1\mu_2 ...\mu_n}\rangle_i^{\rm tree}
\ +\ 
\langle ^Q{\cal O}^{(n),3}_{\mu_1\mu_2 ...\mu_n}\rangle_i^{\rm non-HP}
\ +\ 
\langle ^Q{\cal O}^{(n),3}_{\mu_1\mu_2 ...\mu_n}\rangle_i^{\rm HP}
\ \ \ .
\label{eq:matsum}
\end{eqnarray}
The superscripts on the one-loop contributions indicate
whether or not a hairpin interaction (HP) appears in the diagram.
Of course, there are no HP contributions in QCD.

\subsubsection{Non-Hairpin Contributions}

\begin{figure}[!ht]
\centerline{{\epsfxsize=2.8in \epsfbox{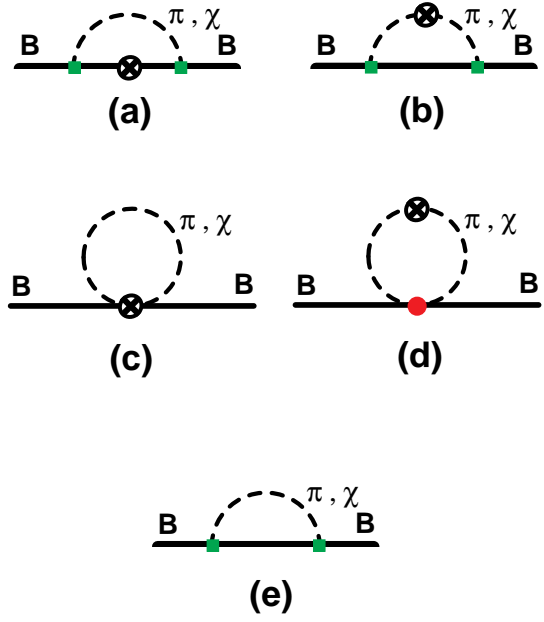}}} 
\vskip 0.15in
\noindent
\caption{\it 
The meson loop diagrams that give leading 
non-analytic contributions
of the form $m_q\log m_q$ to the matrix element of 
$^Q{\cal O}^{(n), 3}_{\mu_1\mu_2\ ... \mu_n}$
between octet baryon states.
The crossed circle denotes an insertion of an operator from
eq.~(\ref{eq:mesontree}) or eq.~(\ref{eq:baryonops}), 
arising directly from the twist-2 operator.
The smaller solid circle denotes an insertion of the
strong two-pion-nucleon 
interaction from the nucleon kinetic energy term
in eq.~(\ref{eq:free}),
while the square denotes an insertion of the axial-vector 
interaction~$\propto F, D, \gamma$.
The label ``$\pi$'' denotes an octet meson,
and ``$\chi$'' denotes an octet fermionic meson.
Diagrams (a)-(d) are vertex corrections while diagram (e) denotes
wavefunction renormalization.
}
\label{fig:Ndiags}
\vskip .2in
\end{figure}

\begin{figure}[!ht]
\centerline{{\epsfxsize=3.2in \epsfbox{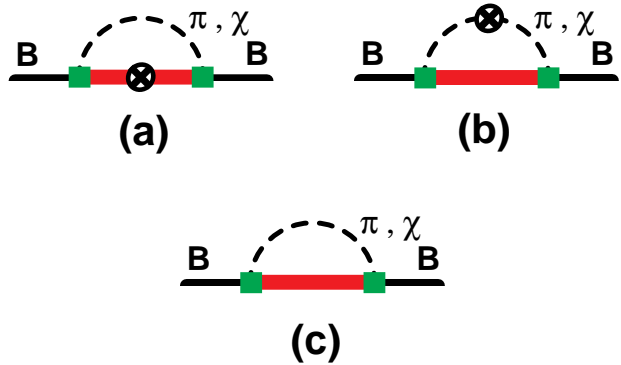}}} 
\vskip 0.15in
\noindent
\caption{\it 
One-loop diagrams with ${\cal T}_{ijk}$
intermediate states that contribute to the 
matrix elements of 
$^Q{\cal O}^{(n), 3}_{\mu_1\mu_2\ ... \mu_n}$
between octet baryon states.
The thick solid line inside the loop denotes a 
${\cal T}_{ijk}$ propagator,
while the dashed line denotes  a meson propagator.
The crossed circle denotes an insertion of an operator from
eq.~(\ref{eq:mesontree}) or eq.~(\ref{eq:baryonops}), 
arising directly from the twist-2 operator, and 
the square denotes an insertion of the strong
interaction vertex~$\propto {\cal C}$.
Diagrams (a) and (b) are vertex corrections while diagram (c) denotes
wavefunction renormalization.
}
\label{fig:deldiags}
\vskip .2in
\end{figure}

The one-loop diagrams in figs.~(\ref{fig:Ndiags}) give 
contributions that are non-analytic in $m_q$
of the form $m_q\log m_q$.  The diagrams in 
figs.~(\ref{fig:deldiags}) also
give terms of the form $m_q\log m_q$
in the limit that the 
decuplet-octet mass splitting, $\Delta$, vanishes, 
but in general give a more complicated
expression.
We write the non-hairpin 
one-loop contribution to the forward matrix element 
in terms of the contributions from the meson mass eigenstates,
\begin{eqnarray}
\langle {\cal O}^{(n),3}_{\mu_1\mu_2 ...\mu_n}\rangle_i^{\rm non-HP}
& = &  
v_{\mu_1} v_{\mu_2}...v_{\mu_n}\ 
{1-\delta^{n1}\over 8\pi^2 f^2}
\left[\ 
\sum_j\ {\cal M}^{(j)}_i\ H_j
\ +\ 
\sum_j\ {\cal N}^{(j)}_i\ J_j
\ \right]
\ -\ {\rm traces}
\\
\langle ^Q{\cal O}^{(n),3}_{\mu_1\mu_2 ...\mu_n}\rangle_i^{\rm non-HP}
& = &  
v_{\mu_1} v_{\mu_2}...v_{\mu_n}\ 
{1-\delta^{n1}\over 8\pi^2 f^2}
\left[\ 
\sum_j\  ^Q{\cal M}^{(j)}_i\ H_j
\ +\ 
\sum_j\  ^Q{\cal N}^{(j)}_i\ J_j
\ \right]
\ -\ {\rm traces}
\ \ \ .
\nonumber
\end{eqnarray}
As we are assuming isospin symmetry in our calculation,
the sum over mesons for QCD corresponds to 
$j=\pi, K, \eta$, while for
QQCD the sum corresponds to 
$j=\pi, K, \eta_s$.
The coefficients resulting from mesons with the mass of the pion,
${\cal M}^{(\pi)}$ and $^Q{\cal M}^{(\pi)}$
[\ ${\cal N}^{(\pi)}$ and $^Q{\cal N}^{(\pi)}$\ ], 
from the diagrams in
fig.~(\ref{fig:Ndiags}) [\ fig.~(\ref{fig:deldiags})\ ]
are tabulated in Table~\ref{tab:pi} [\ Table~\ref{tab:delpi}\ ],
while
the coefficients resulting from mesons with the mass of the kaon,
${\cal M}^{(K)}$ and $^Q{\cal M}^{(K)}$
[\ ${\cal N}^{(K)}$ and $^Q{\cal N}^{(K)}$\ ],
are tabulated in Table~\ref{tab:K} [ Table~\ref{tab:delK}].
\begin{table}[!ht]
\begin{tabular}[h]{ccc}
B  & ${\cal M}^{(\pi)}$ (QCD)
&  $^Q{\cal M}^{(\pi)}$ (QQCD) \\ \hline
$p$ & 
$-\left[1+3(D+F)^2\right] {\cal M}_p^{\rm tree}$
&
$2D(D-3F) {\cal M}_p^{\rm tree}
- \left(\alpha^{(n)}+\beta^{(n)}\right)
2D(D-F)$\\
$\Sigma^+$ & 
$-\left[1+D^2+3F^2\right] {\cal M}_{\Sigma^+}^{\rm tree}
- \left(\alpha^{(n)}-2\beta^{(n)}\right) DF$
& $D^2  {\cal M}_{\Sigma^+}^{\rm tree}
- {1\over 2}  \left(\alpha^{(n)}+2\beta^{(n)}\right)
D^2 $\\
$\Xi^0$ &
$-\left[1+3(D-F)^2\right] {\cal M}_{\Xi^0}^{\rm tree}$
& $0$\\
\end{tabular}
\caption{
Contributions to the matrix elements of
${\cal O}^{(n),3}_{\mu_1\mu_2 ...\mu_n}$
and
$^Q{\cal O}^{(n),3}_{\mu_1\mu_2 ...\mu_n}$
in the octet baryons that are proportional to $H_\pi$.
}
\label{tab:pi}
\end{table}
The pionic contribution to the QCD matrix element in the nucleon 
agrees with the calculations of Refs.~\cite{AS,CJ}, performed
in $SU(2)_L\otimes SU(2)_R$ chiral perturbation theory,
when $D+F\rightarrow g_A$.
\begin{table}[!ht]
\begin{tabular}[h]{ccc}
B  & ${\cal M}^{(K)}$ (QCD)
&  $^Q{\cal M}^{(K)}$ (QQCD) \\ \hline
$p$ & 
$-{1\over 2}\left[1+5D^2+3F^2\right] {\cal M}_p^{\rm tree}
+ \left(\alpha^{(n)}+\beta^{(n)}\right) D(D-F)$
&
$0$\\ \hline
$\Sigma^+$ & 
$-{1\over 2}\left[1+3D^2+3F^2\right] {\cal M}_{\Sigma^+}^{\rm tree}
- {3\over 2}\left(\alpha^{(n)}-2\beta^{(n)}\right) DF$
& $D(D-6F) {\cal M}_{\Sigma^+}^{\rm tree}
+ 4\left(\alpha^{(n)}+\beta^{(n)}\right)DF $
\\
& &$ - {1\over 2}\left(3\alpha^{(n)}+2\beta^{(n)}\right) D^2
 $\\ \hline
$\Xi^0$ &
$-{1\over 2}\left[1+5D^2+3F^2\right] {\cal M}_{\Xi^0}^{\rm tree}
+{3\over 2}\alpha^{(n)} D(D+F)$
& $D(D-6F) {\cal M}_{\Xi^0}^{\rm tree}
+ 2\left(\alpha^{(n)}+\beta^{(n)}\right)DF $\\ & & 
$+{1\over 2}\alpha^{(n)} D^2$
\\
\end{tabular}
\caption{
Contributions to the matrix elements of
${\cal O}^{(n),3}_{\mu_1\mu_2 ...\mu_n}$
and
$^Q{\cal O}^{(n),3}_{\mu_1\mu_2 ...\mu_n}$
in the octet baryons that are proportional to $H_K$.
}
\label{tab:K}
\end{table}
The coefficients resulting from mesons with the mass of the $\eta$,
${\cal M}^{(\eta)}$ [\ ${\cal N}^{(\eta)}$\ ], 
in QCD and the mass of the $\eta_s$,
$^Q{\cal M}^{(\eta_s)}$ [\ $^Q{\cal N}^{(\eta_s)}$\ ], 
in QQCD from the diagrams in
fig.~(\ref{fig:Ndiags}) [\ fig.~(\ref{fig:deldiags})\ ]
are tabulated in Table~\ref{tab:eta} [\ Table~\ref{tab:deleta}\ ].
\begin{table}[!ht]
\begin{tabular}[h]{ccc}
B  & ${\cal M}^{(\eta)}$ (QCD)
&  $^Q{\cal M}^{(\eta_s)}$ (QQCD) \\ \hline
$p$ & 
$0$
&
$0$\\
$\Sigma^+$ & 
$0$ & $0$\\
$\Xi^0$ &
$0$ & 
$-{1\over 3} \left(\alpha^{(n)}-2\beta^{(n)}\right)D^2 $
\end{tabular}
\caption{
Contributions to the matrix elements of
${\cal O}^{(n),3}_{\mu_1\mu_2 ...\mu_n}$
proportional to $H_\eta$ in QCD and
the  matrix elements of
$^Q{\cal O}^{(n),3}_{\mu_1\mu_2 ...\mu_n}$
proportional to $H_{\eta_s}$ in QQCD
in the octet baryons.
}
\label{tab:eta}
\end{table}
\begin{table}[!ht]
\begin{tabular}[h]{ccc}
B  & ${\cal N}^{(\pi)}$ (QCD)
&  $^Q{\cal N}^{(\pi)}$ (QQCD) \\ \hline
$p$ & 
$-{\cal C}^2\left[\ 2 {\cal M}_p^{\rm tree}
+ {10\over 9} \left(\gamma^{(n)}-{\sigma^{(n)}\over 3}\right)\ \right]$
&
$-{\cal C}^2\left[\ {\cal M}_p^{\rm tree}
+ {2\over 3} \left(\gamma^{(n)}-{\sigma^{(n)}\over 3}\right)\ \right]$
\\
$\Sigma^+$ & 
$-{\cal C}^2\left[\ {1\over 3} {\cal M}_{\Sigma^+}^{\rm tree}
+ {1\over 9} \left(\gamma^{(n)}-{\sigma^{(n)}\over 3}\right)\ \right]$
& 
$-{\cal C}^2\left[\ {1\over 6} {\cal M}_{\Sigma^+}^{\rm tree}
+ {1\over 9} \left(\gamma^{(n)}-{\sigma^{(n)}\over 3}\right)\ \right]$
\\
$\Xi^0$ &
$-{\cal C}^2\left[\ {1\over 2} {\cal M}_{\Xi^0}^{\rm tree}
- {1\over 18} \left(\gamma^{(n)}-{\sigma^{(n)}\over 3}\right)\ \right]$
&
$0$
\end{tabular}
\caption{
Contributions to the matrix elements of
${\cal O}^{(n),3}_{\mu_1\mu_2 ...\mu_n}$
and
$^Q{\cal O}^{(n),3}_{\mu_1\mu_2 ...\mu_n}$
in the octet baryons that are proportional to $J_\pi$.
}
\label{tab:delpi}
\end{table}
\begin{table}[!ht]
\begin{tabular}[h]{ccc}
B  & ${\cal N}^{(K)}$ (QCD)
&  $^Q{\cal N}^{(K)}$ (QQCD) \\ \hline
$p$ & 
$-{\cal C}^2\left[\ {1\over 2} {\cal M}_p^{\rm tree}
+ {2\over 9} \left(\gamma^{(n)}-{\sigma^{(n)}\over 3}\right)\ \right]$
&
$0$
\\
$\Sigma^+$ & 
$-{\cal C}^2\left[\ {5\over 3} {\cal M}_{\Sigma^+}^{\rm tree}
+ {11\over 9} \left(\gamma^{(n)}-{\sigma^{(n)}\over 3}\right)\ \right]$
& 
$-{\cal C}^2\left[\ {5\over 6} {\cal M}_{\Sigma^+}^{\rm tree}
+ {13\over 18} \left(\gamma^{(n)}-{\sigma^{(n)}\over 3}\right)\ \right]$
\\
$\Xi^0$ &
$-{\cal C}^2\left[\ {3\over 2} {\cal M}_{\Xi^0}^{\rm tree}
+ {2\over 9} \left(\gamma^{(n)}-{\sigma^{(n)}\over 3}\right)\ \right]$
&
$-{\cal C}^2\left[\ {5\over 6} {\cal M}_{\Xi^0}^{\rm tree}
+ {1\over 9} \left(\gamma^{(n)}-{\sigma^{(n)}\over 3}\right)\ \right]$
\end{tabular}
\caption{
Contributions to the matrix elements of
${\cal O}^{(n),3}_{\mu_1\mu_2 ...\mu_n}$
and
$^Q{\cal O}^{(n),3}_{\mu_1\mu_2 ...\mu_n}$
in the octet baryons that are proportional to $J_K$.
}
\label{tab:delK}
\end{table}
\begin{table}[!ht]
\begin{tabular}[h]{ccc}
B  & ${\cal N}^{(\eta)}$ (QCD)
&  $^Q{\cal N}^{(\eta_s)}$ (QQCD) \\ \hline
$p$ & 
$0$
&
$0$\\
$\Sigma^+$ & 
$-{\cal C}^2\left[\ {1\over 2} {\cal M}_{\Sigma^+}^{\rm tree}
+ {1\over 3} \left(\gamma^{(n)}-{\sigma^{(n)}\over 3}\right)\ \right]$
& 
$0$
\\
$\Xi^0$ &
$-{\cal C}^2\left[\ {1\over 2} {\cal M}_{\Xi^0}^{\rm tree}
+ {3\over 18} \left(\gamma^{(n)}-{\sigma^{(n)}\over 3}\right)\ \right]$
&
$-{\cal C}^2\left[\ {1\over 6} {\cal M}_{\Xi^0}^{\rm tree}
+ {1\over 18} \left(\gamma^{(n)}-{\sigma^{(n)}\over 3}\right)\ \right]$
\end{tabular}
\caption{
Contributions to the matrix elements of
${\cal O}^{(n),3}_{\mu_1\mu_2 ...\mu_n}$
proportional to $J_\eta$ in QCD and
the  matrix elements of
$^Q{\cal O}^{(n),3}_{\mu_1\mu_2 ...\mu_n}$
proportional to $J_{\eta_s}$ in QQCD
in the octet baryons.
}
\label{tab:deleta}
\end{table}
The integrals $H_j$ and $J_j$ are given by 
\begin{eqnarray}
J_j & = & 
\left(m_j^2-2\Delta^2\right)\log\left({m_j^2\over\mu^2}\right)
+2\Delta\sqrt{\Delta^2-m_j^2}
\log\left({\Delta-\sqrt{\Delta^2-m_j^2+ i \epsilon}\over
\Delta+\sqrt{\Delta^2-m_j^2+ i \epsilon}}\right)
\nonumber\\
H_j & = & m_j^2\log\left({m_j^2\over\mu^2}\right)
\ \ \ ,
\label{eq:nonHPints}
\end{eqnarray}
where $\Delta$ is the decuplet-octet mass splitting and $m_j$ is the mass of
the meson, and we have retained only those terms that are non-analytic in $m_j$
or are required for the correct decoupling in the $\Delta\rightarrow\infty$
limit.  It is clear that for $\Delta=0$, $J_j=H_j$.

\subsubsection{Hairpin Contributions}

There are contributions to some 
matrix elements from
one-loop diagrams involving the hairpin interactions in QQCD.
\begin{figure}[!ht]
\centerline{{\epsfxsize=3.2in \epsfbox{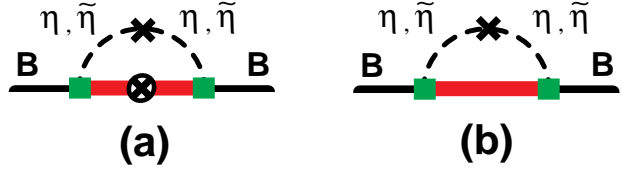}}} 
\vskip 0.15in
\noindent
\caption{\it 
One-loop hairpin diagrams with ${\cal T}_{ijk}$
intermediate states that contribute to the 
matrix elements of 
$^Q{\cal O}^{(n), 3}_{\mu_1\mu_2\ ... \mu_n}$
between octet baryon states.
The thick solid line inside the loop denotes a 
${\cal T}_{ijk}$ propagator,
while the dashed line denotes  an 
$\eta=\eta_{u,d,s}$ or $\tilde\eta=\tilde\eta_{u,d,s}$ 
propagator.
The crossed circle denotes an insertion of an operator from
eq.~(\ref{eq:baryonops}), 
arising directly from the twist-2 operator, and 
the square denotes an insertion of the strong
interaction vertex~$\propto {\cal C}$.
The cross on the meson propagator denotes an insertion
of a hairpin interaction with coefficient $M_0^2$ or 
$\alpha_\Phi$.
Diagram (a) is a vertex correction while diagram (b) denotes
wavefunction renormalization.
}
\label{fig:HPdiags}
\vskip .2in
\end{figure}
Explicit computation of the diagrams in fig.~\ref{fig:HPdiags} yields
\begin{eqnarray}
\langle ^Q{\cal O}^{(n),3}_{\mu_1\mu_2 ...\mu_n}\rangle_i^{\rm HP}
& = &   - v_{\mu_1} v_{\mu_2}...v_{\mu_n}\ 
{1\over 16\pi^2 f^2}\  {2 {\cal C}^2\over 9}\ h_i\ 
\left[\ M_0^2\left(
\overline{I}^{\Delta\Delta}_{ss}+\overline{I}^{\Delta\Delta}_{uu}-2
\overline{I}^{\Delta\Delta}_{us}\right)
\right.\nonumber\\
& & \left.
\ -\ 
\alpha_\Phi \left( m_{\eta_s}^2\left(
\overline{I}^{\Delta\Delta}_{ss} - \overline{I}^{\Delta\Delta}_{us}
\right)
+ 
m_{\eta_u}^2\left(
\overline{I}^{\Delta\Delta}_{uu} - \overline{I}^{\Delta\Delta}_{us}
\right)\right)
\right]
\ -\ {\rm traces}
\ \ \ ,
\label{eq:hps}
\end{eqnarray}
where $\alpha_\Phi$ and $ M_0$ are defined in eq.~(\ref{eq:lagpi})
and ${\cal C}$ is the axial coupling between the decuplet and octet.
The integrals $\overline{I}^{\Delta\Delta}_{ab}$ are defined to 
be~\cite{Sav01}
$\overline{I}_{q q^\prime}^{\Delta\Delta}~=~\overline{I}(m_{\eta_q}, 
m_{\eta_{q^\prime}}, \Delta, \Delta, \mu)$,
where
\begin{eqnarray}
& & \overline{I}(m_1, m_2, \Delta_1, \Delta_2, \mu) 
=  
{
\left[ \ Y(m_1,\Delta_1, \mu)+ Y(m_2,\Delta_2, \mu)
- Y(m_1,\Delta_2, \mu)-Y(m_2,\Delta_1, \mu)
\ \right]
\over
[\Delta_1-\Delta_2][m_1^2-m_2^2]}
\ \ \ ,
\label{eq:Ibardef}
\end{eqnarray}
with
\begin{eqnarray}
Y(m,\Delta, \mu) & = & 
\left[m^2-{2\over 3}\Delta^2\right]\Delta\log \left({m^2\over\mu^2}\right)
\ +\ 
{2\over 3}\left[\Delta^2-m^2\right]^{3\over 2}
\log\left({\Delta-\sqrt{\Delta^2-m^2+i\epsilon}\over
\Delta+\sqrt{\Delta^2-m^2+i\epsilon}}\right)
\ .
\end{eqnarray}
The Clebsch-Gordan coefficients, $h_i$, 
appearing in eq.~(\ref{eq:hps})
are given in Table~\ref{tab:HP}.
It is important to note that there is no contribution from one-loop hairpin
diagrams with ${\cal B}_{ijk}$ (containing the octet)
intermediate states, since  the 
vertex diagrams are exactly canceled by wavefunction
renormalization.
\begin{table}[!ht]
\begin{tabular}[h]{cccc} 
& B & $h_i$ & \\ \hline
& $p$ &\ \ $0$& \\
& $\Sigma^+$ &\ \ ${1\over 6}\left(5\alpha^{(n)}+2\beta^{(n)}
+4\gamma^{(n)}-{4\over 3}\sigma^{(n)}
\right)$& \\
& $\Xi^0$ &\ \ ${1\over 6}\left(\alpha^{(n)}+4\beta^{(n)}
+2\gamma^{(n)}-{2\over 3}\sigma^{(n)}
\right)$&
\end{tabular}
\caption{
Contributions to the matrix elements of the isovector twist-2 operators
$^Q{\cal O}^{(n),3}_{\mu_1\mu_2 ...\mu_n}$
from one-loop hairpin diagrams involving ${\cal T}_{ijk}$ intermediate 
states.
}
\label{tab:HP}
\end{table}
Further, the hairpin diagrams do not contribute 
to the matrix elements in the nucleon in the
limit of exact isospin symmetry, as the possible intermediate states in 
${\cal T}_{ijk}$ are inaccessible.

\subsection{Matrix Elements in the Nucleon}

Our motivation for undertaking this work is to construct a framework in which
experimental data can be compared with the predictions of lattice 
QCD.  As data exists
only for the nucleon, it is natural to ask why we bothered with the other
members of the baryon octet.
As the present values of $m_q^{\rm latt.}$ are large, and in fact 
$m_s^{\rm latt.}\sim m_{u,d}^{\rm latt.}$, we
consider three light quarks as opposed to just two.
Thus the quantities $\alpha^{(n)},\beta^{(n)}, \gamma^{(n)}$,
$\sigma^{(n)}$ 
and combinations of the $b_i^{(n)}$
must all be determined, in addition to the strong interaction
couplings $F, D$ and ${\cal C}$ in order to perform the 
quark mass extrapolation at one-loop order.
For a one-loop analysis as we have presented here,
the coefficients
$\gamma^{(n)}$ and $\sigma^{(n)}$ can be determined at tree-level from 
calculations of matrix elements in the baryon decuplet, and it is consistent 
to use these values in the one-loop expression for the octet baryons.
Calculation of the proton, $\Sigma^+$ and $\Xi^0$ matrix elements will allow
for a determination of $\alpha^{(n)}$, $\beta^{(n)}$, 
and combinations of the $b_i^{(n)}$,
and hence the 
$m_q$-dependence of the nucleon matrix element.
While the hairpin interactions, $M_0^2$ and $\alpha_\Phi$, do not contribute
directly to the matrix elements in the nucleon, they do contribute to the
matrix elements in the other octet baryons, and so will indirectly
influence the calculation of the matrix elements in the nucleon.

\begin{figure}[!ht]
\centerline{{\epsfxsize=4.0in \epsfbox{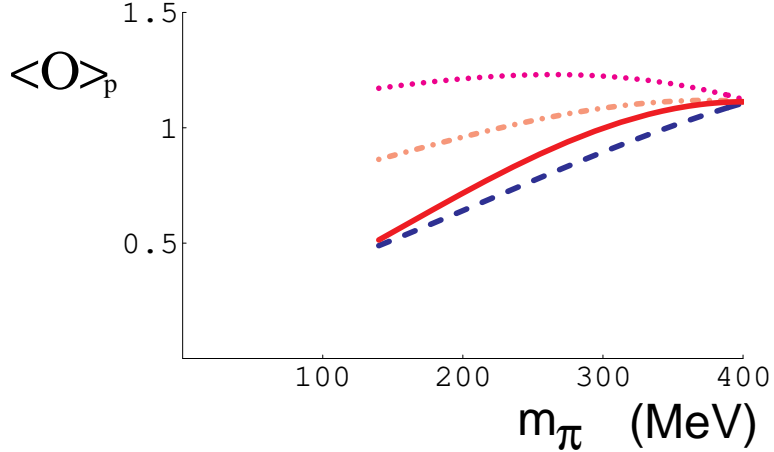}}} 
\vskip 0.15in
\noindent
\caption{\it 
The variation of the matrix element of the 
isovector twist-2 operator in the proton.
The kaon mass is fixed to its experimental value.
The solid curve corresponds to a QCD matrix element with
$\gamma^{(n)}=\sigma^{(n)}=0$, $\alpha^{(n)}=+2$,
$\beta^{(n)}=+1$, $b_i^{(n)}=0$,
$F=0.5$, $D=0.8$, ${\cal C}=1.8$
and $\Delta~=~293~{\rm MeV}$.
The dot-dashed curve corresponds to the QQCD matrix element with 
the parameters of set (a) described in the text,
$\gamma^{(n)}=\sigma^{(n)}=0$, 
$\alpha^{(n)}=+2.35$, $\beta^{(n)}=+1$, $b_i^{(n)}=0$,
$F=0.5$, $D=0.8$, ${\cal C}=1.8$, and $\Delta=293~{\rm MeV}$.
The dotted curve corresponds to the QQCD matrix element with 
the parameters of set (b) described in the text,
$\gamma^{(n)}=\sigma^{(n)}=0$, 
$\alpha^{(n)}=+2.11$, $\beta^{(n)}=+1$, $b_i^{(n)}=0$,
$F=0.5$, $D=0.8$, ${\cal C}=1.8$, and $\Delta=500~{\rm MeV}$.
Finally, the dashed curve corresponds to the QQCD matrix element with 
the parameters of set (c) described in the text,
$\gamma^{(n)}=\sigma^{(n)}=0$, 
$\alpha^{(n)}=+0.89$, $\beta^{(n)}=+1$, $b_i^{(n)}=0$,
$F=2.0$, $D=2.0$, ${\cal C}=1.0$, and 
$\Delta=500~{\rm MeV}$.
}
\label{fig:T2}
\vskip .2in
\end{figure}

To get a feel for the 
$m_q$-dependence of the matrix elements in the proton, 
we pick some values for the parameters describing the 
proton matrix element and vary the pion mass.
For the QCD matrix element shown in fig.~\ref{fig:T2}, 
we choose $\gamma^{(n)}=\sigma^{(n)}=0$, and 
$\alpha^{(n)}=+2$ , $\beta^{(n)}=+1$ and $b_i^{(n)}=0$
for $n\ne 1$
(this does not eliminate contributions from the ${\cal T}_{ijk}$ 
intermediate states, as they contribute in wavefunction renormalization).
Also, we use the tree-level QCD values for the axial coupling constants
$F=0.5$, $D=0.8$~\cite{FMJM98} and ${\cal C}=1.8$~\cite{BSS92},
and the measured value of the decuplet-octet mass splitting, 
$\Delta=293~{\rm MeV}$.
For the QQCD matrix elements shown in fig.~\ref{fig:T2}, we choose three
different parameter sets, set (a), set (b) and set (c), 
that reproduce the QCD matrix element at
$m_\pi=400~{\rm MeV}$.
Set (a) has $\gamma^{(n)}=\sigma^{(n)}=0$, 
$\alpha^{(n)}=+2.35$, $\beta^{(n)}=+1$ and $b_i^{(n)}=0$
for the tree-level matrix elements,
with $F=0.5$, $D=0.8$, ${\cal C}=1.8$, and $\Delta=293~{\rm MeV}$.
Set (b) has $\gamma^{(n)}=\sigma^{(n)}=0$, 
$\alpha^{(n)}=+2.11$, $\beta^{(n)}=+1$ and $b_i^{(n)}=0$
for the tree-level matrix elements,
with $F=0.5$, $D=0.8$, ${\cal C}=1.8$, and $\Delta=500~{\rm MeV}$~\footnote{
Analysis of 
quenched simulations suggests that the  mass splitting $\Delta$ 
is $\sim 400-500~{\rm MeV}$~\cite{YLTW}.
}.
In contrast, set (c) has 
$\gamma^{(n)}=\sigma^{(n)}=0$, 
$\alpha^{(n)}=+0.89$, $\beta^{(n)}=+1$ and $b_i^{(n)}=0$
for the tree-level matrix elements,
with 
$F=2.0$, $D=2.0$, ${\cal C}=1.0$, and $\Delta=500~{\rm MeV}$.
%

\section{Conclusions}

It is important to know the parton distribution
functions of the nucleon in QQCD.
In order to obtain these predictions from the 
lattice computations of the
foreseeable future, extrapolations in $m_q$ from the lattice values
down to the physical values will still be required.
We have taken the first step toward establishing the 
theoretical $m_q$-dependence by  computing the leading 
$m_q$-dependence of the 
forward matrix element of the isovector twist-2 operators
in the octet baryons in QQCD using Q$\chi$PT.  
It is certainly true that higher order
calculations are required, not only to establish the convergence of the 
expansion about the chiral limit, but also to allow an extrapolation
to the relatively large values of the lattice quark masses.
Even in QCD, the convergence of $SU(3)$ chiral perturbation theory varies
from observable to observable due to the large value of the kaon mass.
Recently, it has been observed that 
baryon masses obtained from QQCD simulations by
applying the QCD extrapolation 
agree reasonably well with QCD simulations of the masses~\cite{YLTW}.
It will be a pleasant surprise if this empirical feature is
found to hold for other observables, and particularly the matrix elements of
the twist-2 operators. 
We look forward to  a complete analysis of the quenched lattice data, 
including an extraction of the constants that describe these matrix
elements and  eventual predictions at the physical value of $m_\pi$.

\vskip 0.5in

MJS is supported in
part by the U.S. Dept. of Energy under Grant No.  DE-FG03-97ER4014
and JWC is supported in
part by the U.S. Dept. of Energy under Grant No. DE-FG02-93ER-40762.


\begin{references}

\bibitem{latticeQCD}
G.~Martinelli and C.T.~Sachrajda,
{\it Phys. Lett.} {\bf B 196}, 184 (1987);
{\it Nucl. Phys.} {\bf B306}, 865 (1988);
M.~G\"ockeler {\em et al.},
{\it Phys. Rev.} {\bf D 53}, 2317 (1996);
M.~G\"ockeler {\em et al.},
{\it Nucl. Phys. Proc. Suppl.} {\bf 53}, 81 (1997);
C.~Best {\it et al.},
{\tt hep-ph/9706502}.

\bibitem{latticeQCDu}
D.~Dolgov {\em et al.},
{\it Nucl. Phys. Proc. Suppl.} {\bf 94}, 303 (2001);
D.~Dolgov,
Ph.D. thesis, MIT, Sep. 2000.

\bibitem{Negele}
J. W. Negele,
{\it Nucl. Phys.} {\bf A699}, 18 (2002). 

\bibitem{Pqqcd}
S.~R.~Sharpe and N.~Shoresh,
{\tt hep-lat/0108003};
{\tt hep-lat/0011089};
{\it Phys. Rev.} {\bf D62}, 094503 (2000);
{\it Nucl. Phys. Proc. Suppl.} {\bf 83}, 968 (2000);
S.~R.~Sharpe,
{\it Phys. Rev.} {\bf D56}, 7052 (1997);
C.~W.~Bernard and M.~F.~L.~Golterman,
{\it Phys. Rev.} {\bf D49}, 486 (1994).

\bibitem{mqothers}
A. W. Thomas, W. Melnitchouk and F. M. Steffens,
{\it Phys. Rev. Lett.} {\bf 85}, 2892 (2000);
W. Detmold,  W. Melnitchouk, J. W. Negele, D. B. Renner 
and A. W. Thomas,
{\tt hep-lat/0103006};
W. Detmold,  D. B. Leinweber, W. Melnitchouk, A. W. Thomas,
and S. V. Wright,
{\tt nucl-th/0104043}.


\bibitem{AS}
D. Arndt and M.J. Savage, 
{\it Nucl. Phys.} {\bf A697}, 429 (2002).

\bibitem{CJ}
J-.W. Chen and X. Ji,
{\tt hep-ph/0105197}.

\bibitem{largeN}
J-.W. Chen and X. Ji,
{\tt hep-ph/0105296}.

\bibitem{conv}
J-.W. Chen and X. Ji,
{\it Phys. Rev. Lett.} {\bf 87}, 152002 (2001).

\bibitem{Sharpe90}
S. R. Sharpe,
{\it Nucl. Phys.}  {\bf B17} (Proc. Suppl.), 146 (1990).

\bibitem{S92}
S. R. Sharpe, 
{\it Phys. Rev.} {\bf D46}, 3146 (1992). 

\bibitem{BG92}
C. Bernard and M. F. L. Golterman,
{\it Phys. Rev.}  {\bf D46}, 853 (1992).

\bibitem{LS96}
J. N. Labrenz and S. R. Sharpe,
{\it Phys. Rev.}  {\bf D54}, 4595 (1996).

\bibitem{Sav01}
M. J. Savage,
{\tt nucl-th/0107038}.

\bibitem{S91}
S. R. Sharpe, 
{\it Phys. Rev.} {\bf D41}, 3233 (1990); 

\bibitem{Pall97}
G. Colangelo and  E Pallante,
{\it Nucl. Phys.} {\bf B520}, 433 (1998).

\bibitem{G94}
M. F. L.  Golterman,
{\it Acta Phys. Polon.} {\bf B25}, 1731 (1994).

\bibitem{JMheavy}
E. Jenkins and  A. V. Manohar,
{\it Phys. Lett.} {\bf B255}, 558 (1991). 

\bibitem{JMaxial}
E. Jenkins and  A. V. Manohar,
{\it Phys. Lett.} {\bf B259}, 353 (1991). 

\bibitem{Jmass}
E. Jenkins,
{\it Nucl. Phys.} {\bf B368}, 190 (1992). 

\bibitem{chiralN}
E. Jenkins and A. V. Manohar,
{\it Baryon Chiral Perturbation Theory},
talks presented at the workshop on {\it Effective Field Theories
of the Standard Model}, Dobogoko, Hungary (1991);

\bibitem{chiralUlf}
For a recent review see
U.-G. Mei\ss ner,
Essay for the Festschrift in honor of Boris Ioffe, 
in ``Encyclopedia of Analytic QCD'', 
edited by M. Shifman, to be published by World Scientific. 
{\tt hep-ph/0007092}.

\bibitem{GP}
M. F. L. Golterman and E. Pallante,
{\it JHEP} 0110, 037 (2001);
{\tt hep-lat/0108029};
{\tt hep-lat/0110183};

\bibitem{CSPQ}
J.-W. Chen and M. J. Savage,
{\tt hep-lat/0111050}.

\bibitem{FMJM98}
R. Flores-Mendieta, E. Jenkins and A. V. Manohar,
{\it Phys. Rev.} {\bf D58}, 094028 (1998).

\bibitem{BSS92}
M. N. Butler, M. J. Savage and R. P. Springer,
{\it Nucl. Phys.} {\bf B399}, 69 (1993).

\bibitem{YLTW}
R. D. Young, D. B. Leinweber, A. W. Thomas and S. V. Wright,
{\tt hep-latt/0111041}.

\end{references}
\end{document}